\documentclass[12pt,a4paper]{article}
\usepackage[utf8]{inputenc}
\usepackage[T1]{fontenc}
\usepackage{amsmath}
\usepackage{amsfonts}
\usepackage{amssymb}
\usepackage{authblk}
\usepackage{graphicx}
\date{}
\begin{document}
\title{Interacting cosmic strings and Dark matter-For the case of missing stars}
\author[1,2]{Natarajan Shriethar \thanks{natarajan@spacequanta.com}}
\author[1,2]{Nageswaran Rajendran \thanks{eswar.quanta@gmail.com}}
\affil[1]{SpaceQuanta Lab, Devakottai, Tamilnadu, India}
\affil[2]{Qubitor lab, Singapore}


\maketitle

\begin{abstract}

This work discusses a few theories including the interaction of dark matter, cosmic strings, and locally coupled dark energy. The paper also examines mathematical models used to describe the pressure and density within a star, including the polytropic relationship and the Lane-Emden equation. Simulation results from the IllustrisTNG datasets are also presented, providing insights into the interacting dark matter solutions.   With the derived solutions this paper, it explores the possible causes for the sudden disappearance of the star PHL293B-LBV.
\end{abstract}

\section{Introduction}

 Cosmic strings are theoretical objects proposed to be extremely dense, thin, and massive, and are thought to have formed shortly after the Big Bang \cite{vilenkin1994cosmic}. 
 
 Dark matter is understood as a critical component that consists significant portion of the universe's total mass-energy component. Still, it does not interact with light, making it invisible to telescopes.

 The interaction of dark matter with matter is proposed in \cite{barkana2018possible}. Such interactions can be conformed with 21-cm transition lines.  But on the observational side, detection of such dark matter is an uphill task. Dark matter would penetrate every stellar object without providing any traces of it. Few theories suggest that neutron stars are a trapping place for dark matter. It is also suggested that the dark matter can be clumped together gravitationally \cite{loeb2005small}. 
 
  
 

  The present work suggests that cosmic strings could be made up of some form of dark matter and that their motion and evolution can affect the distribution of matter in the universe, potentially leading to the formation of voids. Additionally, the possible interaction of dark matter with spherically symmetric stellar objects such as stars is also discussed. The present work  proposes the interaction of dark matter with cosmic strings, as a result, the spherically symmetric stellar objects can have significant effects on the observations and properties of these objects. The role of dark energy in the local stellar object is also considered for the solution.
 
 The obtained solutions are applied to the mysterious escape of the star.  To verify the solutions obtained, the SageManifolds package is employed \cite{gourgoulhon2015tensor}. The SageManifolds package is applied for the calculation of various cosmological solutions \cite{natarajan2021conformal,shriethar2020conformal}.

As an application of the interactive solutions, we try to understand the causes of the mystery of a massive star from the observations.
The massive star PHL293B-LBV \cite{izotov2011vlt} (luminous blue variable) has disappeared from observation since 2019. The mysterious part is that there is no direct consequence, such as a supernova explosion that could be observed after the disappearance.
 Using the very large telescope from the European Southern Observatory, it has been found that a massive star in the Kinman Dwarf galaxy PHL 293B has disappeared from the observations very suddenly \cite{core1}. Even though the Espresso instrument attempted to point out the signs of the disappearance, there was also no evidence that could be found. Similarly, the X-Shooter instrument also provided no trace. Hence we attempt to provide a possible explanation for the sudden disappearance.

 By running large-scale simulations of the universe, IllustrisTNG \cite{nelson2019illustristng,pillepich2018first,springel2018first} is able to generate lots of predictions for the properties and distribution of galaxies, to test our current understanding of the universe. The IllustrisTNG simulations use a large number of particles to represent both dark matter and baryonic (normal) matter and use mathematical equations to model the physical processes that govern the behavior of these particles, including gravity, hydrodynamics, radiative cooling and heating, star formation and feedback, and supermassive black hole growth. The simulations are run on some of the most powerful supercomputers available and are able to model large volumes of the universe with high resolution. In this paper, we will use IllustrisTNG simulation data sets to test the hypothesis that the local domination of the dark energy, and to explore the sudden disappearance of the star PHL293B-LBV.
 
The present work also examines the mass-to-radius relationship of stellar objects and how the addition of dark matter may affect this relationship, potentially leading to the sudden disappearance of a star. This idea is presented as a possible explanation for the dimming of the star Betelgeuse.

 

\section{The Cosmic strings}\label{cosi}

Cosmic strings are hypothetical topological defects that are thought to have formed during the early universe, potentially as a result of phase transitions in the fundamental fields that governed the universe at that time. They are one-dimensional objects, or "strings," that can stretch across the entire universe and have a huge mass per unit length.

In the complex scalar field, $\phi$ the potential energy function is written with an undetermined phase and is defined as

\begin{equation}\label{csa1}
V(\phi) = \left(|\phi|^2 - \eta_v^2\right)
\end{equation}

 This means that the phase is not fixed, and can take on any value, while the magnitude of the complex scalar field is fixed by the value of $\eta_v^2$.

The minimal energy configuration is defined as

\begin{equation}\label{csa2}
|\phi| = \eta_v
\end{equation}

The equation \ref{csa2} suggests that the minimal energy configuration is a specific state that the complex scalar field can take on, characterized by a specific value of its magnitude. It provides a mathematical representation of the conditions that the complex scalar field must meet to have the lowest possible energy in space-time.

 The minimal energy configuration of the complex scalar field determines the properties of the cosmic string that is formed, such as its mass and tension.

In manifold $\mathcal{M}$ the vacua can form cosmic voids. The closed loop around the circle can not be shrunk into a point. But it can be warped around a circle on a vacuum in $n=\pm 1, \pm 2,\ldots$ times. There may be $n$ possible string loops around a closed path in physical space. At the center of the string, $\phi = 0$.

These voids are the locations where the complex scalar field has a value of zero.

The Lagrangean on the straight string is obtained as

\begin{equation}\label{csa3}
L = \left |\partial_\mu \Phi \right |^2 - \frac{\lambda}{4}(|\Phi|^2-\eta^2)^2
\end{equation}

With $\lambda$ dimensionless coupling constant, $\eta$ is the vacuum expectation of the field $\phi$. The combination of these two terms in the Lagrangian function describes the dynamics of the string, how it moves, and how it interacts with the space-time around it.

In the equation \ref{cs3} the variable $\lambda$ is a dimensionless coupling constant, which is a numerical value that is used to adjust the strength of the interactions between the field and the space-time. And $\eta$ is the vacuum expectation of the field, which is a specific value that the field can take on and represents the lowest energy state of the field.

 The static sting solutions are obtained as

\begin{equation}\label{csa4}
\phi (x,y) = \eta f (m,\rho) e^{i n \theta}
\end{equation}

$(\rho,\theta)$ are the polar coordinate in $x,y$ plane.
 The equation \ref{cs4} describes the static string solutions  for the complex scalar field, represented by the variable $\phi$. 
 
 The equation defines the complex scalar field as a function of two variables: $x$ and $y$, represented by the symbol $\phi(x,y)$.

 It concludes, 

\begin{equation}\label{csa5}
m^2 = \lambda \eta^2
\end{equation}

It states that the square of the variable $m$ is equal to the product of the dimensionless coupling constant $\lambda$ and the square of the vacuum expectation of the field $\eta$. The equation \ref{csa5} suggests that the value of $m$ is determined by the properties of the complex scalar field and the space-time around it, represented by the variables $\lambda$ and $\eta$.

The winding number $n$ is an integer that describes the number of times the closed loop of the cosmic string is wrapped around a circle. The winding number is a crucial concept in understanding the properties of cosmic strings, as it determines how the string interacts with the space-time around it.

 The energy density of a cosmic string is defined as 

\begin{equation}\label{csa6}
{\cal E} = |\vec{\nabla} \Phi|^2 + V(\Phi)
\end{equation}

The energy density in equation \ref{csa6} is packed within $\rho \sim m $  and that decreases as $m \to \frac{1}{\rho}$. 

Equation \ref{csa6} defines the energy density of a cosmic string, where $\mathcal{E}$ is the energy density, $|\vec{\nabla} \Phi|^2$ is the kinetic energy density, which is a measure of how much the field is changing at a point, and $V(\Phi)$ is the potential energy density of the field.

Additionally, the energy density decreases as $m \to \frac{1}{\rho}$, which means that as the winding number gets very large the energy density decreases and approaches $\frac{1}{\rho}$ in the limit of large $m$.

The metric of the static cosmic string is written as \cite{van2013geometry}
\begin{equation}\label{cs1}
dr^2 = -dt^2+dr^2+\left(1-\frac{\mu}{2 \pi}\right)^2 r^2 d \theta^2 + dz^2
\end{equation}

The connection terms are calculated for the static string metric

\begin{eqnarray}\label{cs2}
\Gamma_{ \phantom{\, r} \, {\theta} \, {\theta} }^{ \, r \phantom{\, {\theta}} \phantom{\, {\theta}} } & = & -\frac{4 \, \pi^{2} r - 4 \, \pi r \mu\left(r\right) + r \mu\left(r\right)^{2} - {\left(2 \, \pi r^{2} - r^{2} \mu\left(r\right)\right)} \frac{\partial\,\mu}{\partial r}}{4 \, \pi^{2}} \\ \Gamma_{ \phantom{\, {\theta}} \, r \, {\theta} }^{ \, {\theta} \phantom{\, r} \phantom{\, {\theta}} } & = & \frac{2 \, \pi - r \frac{\partial\,\mu}{\partial r} - \mu\left(r\right)}{2 \, \pi r - r \mu\left(r\right)} 
\end{eqnarray}
The right-hand side of each equation gives the expression for the Christoffel symbol in terms of the metric tensor, which in this case is the static string metric of the cosmic string, and the partial derivatives of the metric with respect to the coordinates.
The Killing vector is satisfied for the corresponding cosmic strings. It  means that the cosmic strings possess symmetries and therefore it has Killing vectors.

The Walker-Penrose Killing vector is written in the form

\begin{equation}\label{cs3}
K = \rho^2 (\underline{\ell}\otimes \underline{k} + (\underline{k}\otimes \underline{\ell}) + r^2 g
\end{equation}
The Walker-Penrose Killing vector is a specific type of Killing vector that is defined for spacetimes that have a null Killing vector.  It is used as a tool for studying the properties of spacetimes that have a null Killing vector.
Then the Walker-Penrose Killing vectors are obtained as

\begin{eqnarray}\label{cs4}
K_{ \, t \, t }^{ \phantom{\, t}\phantom{\, t} } & = & \frac{a^{4} + a^{2} r^{2} + 2 \, m r^{3}}{a^{2} - 2 \, m r + r^{2}} \\ K_{ \, t \, {\phi} }^{ \phantom{\, t}\phantom{\, {\phi}} } & = & -\frac{a^{3} + a r^{2}}{a^{2} - 2 \, m r + r^{2}} \\ K_{ \, r \, r }^{ \phantom{\, r}\phantom{\, r} } & = & -a^{2} + 2 \, m r \\ K_{ \, {\theta} \, {\theta} }^{ \phantom{\, {\theta}}\phantom{\, {\theta}} } & = & \frac{{\left(4 \, \pi^{2} - 4 \, \pi \mu + \mu^{2}\right)} r^{4}}{4 \, \pi^{2}} \\ K_{ \, {\phi} \, t }^{ \phantom{\, {\phi}}\phantom{\, t} } & = & -\frac{a^{3} + a r^{2}}{a^{2} - 2 \, m r + r^{2}} \\ K_{ \, {\phi} \, {\phi} }^{ \phantom{\, {\phi}}\phantom{\, {\phi}} } & = & \frac{a^{2} r^{2} - 2 \, m r^{3} + r^{4} + a^{2}}{a^{2} - 2 \, m r + r^{2}} \end{eqnarray}

This is a set of equations that defines the Walker-Penrose Killing vectors for a cosmic string. Each equation represents a different component of the Walker-Penrose Killing vector, $K_{ , t , t }, K_{ , t , {\phi} }, K_{ , r , r }, K_{ , {\theta} , {\theta} }, K_{ , {\phi} , t }, K_{ , {\phi} , {\phi} }$. Each component is a function of variables such as $a$, $m$, $r$, and $t$ and they are defined in terms of these variables and their functions of them.

\begin{equation}\label{cs5}
\begin{split}
k = \left( \frac{a^{2} + r^{2}}{2 \, {\left(a^{2} \cos\left({\theta}\right)^{2} + r^{2}\right)}} \right) \frac{\partial}{\partial t } + \left( -\frac{a^{2} - 2 \, m r + r^{2}}{2 \, {\left(a^{2} \cos\left({\theta}\right)^{2} + r^{2}\right)}} \right) \frac{\partial}{\partial r } \\ + \frac{a}{2 \, {\left(a^{2} \cos\left({\theta}\right)^{2} + r^{2}\right)}} \frac{\partial}{\partial {\phi} }
\end{split}
\end{equation}
Equation \ref{cs5} is an equation defining a Killing vector field named as $k$ in terms of coordinates $t,r$ and $\theta$, which generate a symmetry of the given spacetime.
The Killing vector is a vector field that generates an infinitesimal transformation that preserves the metric of the spacetime. 
This Killing vector is defined in a specific coordinate system, a 2+1 decomposition of the four-dimensional spacetime metric.

\begin{equation}\label{cs6}
\ell = \left( \frac{a^{2} + r^{2}}{a^{2} - 2 \, m r + r^{2}} \right) \frac{\partial}{\partial t } +\frac{\partial}{\partial r } + \left( \frac{a}{a^{2} - 2 \, m r + r^{2}} \right) \frac{\partial}{\partial {\phi} }
\end{equation}

Similar to the previous equation \ref{cs5}, the equation \ref{cs6} defines a vector field that generates an infinitesimal transformation that preserves the metric of the spacetime.

\begin{equation}\label{cs7}
\rho = r^2 + (a \cos \theta)^2
\end{equation}

In equation \ref{cs7} $\rho$ is the radial distance from the origin in the two-dimensional plane. It is defined as the sum of the square of the radial distance from the origin, $r^2$, and the square of the distance from the origin along the x-axis, $(a \cos \theta)^2$.
$\rho$ is often used as a radial coordinate in cylindrical and polar coordinates.

Equations (\ref{cs3}), (\ref{cs4}), (\ref{cs5}), (\ref{cs6}) and (\ref{cs7}) collectively represent equations that are used to describe properties of cosmic strings. These equations describe the different physical and mathematical characteristics of cosmic strings.

The Kretschmann scalar, also known as the curvature scalar, is a scalar quantity used in general relativity to measure the curvature of spacetime. It is defined as the contraction of the Riemann tensor $(R_{abcd})$ with itself, that is $K = R_{abcd} R^{abcd}$, where Riemann tensor is a mathematical object that describes the curvature of spacetime.

The Kertsmann Scalar is obtained for the cosmic strings as

\begin{equation}\label{cs8}
K = R_{abcd} R^{abcd}=\frac{4 \, {\left(r^{2} \frac{\partial^{2}}{(\partial r)^{2}}\mu\left(r\right)^{2} + 4 \, r \frac{\partial}{\partial r}\mu\left(r\right) \frac{\partial^{2}}{(\partial r)^{2}}\mu\left(r\right) + 4 \, \frac{\partial}{\partial r}\mu\left(r\right)^{2}\right)}}{4 \, \pi^{2} r^{2} - 4 \, \pi r^{2} \mu\left(r\right) + r^{2} \mu\left(r\right)^{2}}
\end{equation}

The equation (\ref{cs8}) shows the Kretschmann scalar for the cosmic strings. It defines the Kretschmann scalar as a function of the variable $\mu(r)$ and $r$, where $\mu(r)$ denotes the mass per unit length of the cosmic string, and $r$ denotes the radial distance from the origin. The equation also includes partial derivatives of $\mu(r)$, in terms of the square of the function and the product of the function and its derivatives with respect to $r$.

Thus the non-vanishing Kertsmann scalar of the static cosmic string is observed. It can perturb the local stellar object. The interaction of the cosmic string with the stellar object will eventually create voids. Those voids will allow no observation of the stellar object. Even though that phenomenon may not impact every stellar object, the more possible consequence may be on the stellar objects with certain special properties with additional dark matter or the interaction of the local dark energy.

Theories suggest that as cosmic strings move through the universe they would act as "seeds" around which matter would clump together, forming galaxy clusters and large-scale structures. But, they would also create "voids" in the universe, large empty regions with little or no matter.

 Some theories propose that as the cosmic strings move through space they would pull matter behind them, creating a "wake" of low-density regions in their wake \cite{fernandez2020cosmic,sornborger1997structure}. Other theories propose that the cosmic strings themselves would act as barriers, preventing the flow of matter across them and creating low-density regions on either side of the string \cite{pismen1999vortices}. Cosmic strings, being highly dense objects with a large amount of mass per unit length, are a possible candidate for dark matter. Some theories suggest that cosmic strings could be made up of some form of dark matter \cite{cui2009nonthermal}. Other theories propose the existence of particle-like forms of dark matter like Weakly Interacting Massive Particles (WIMPs) and axions, which are also possible candidates for dark matter. Dark matter  could have an effect on the distribution of matter in the universe, and thus affect the formation of voids.
  
 \section{The dark matter}\label{dark1}
 
  Dark matter particles are believed to be weakly interacting massive particles (WIMPs) that do not interact with light or other forms of electromagnetic radiation. As a result, they are extremely difficult to detect directly. However, it is believed that dark matter particles can interact with normal matter through weak nuclear force.
 
 As the dimming of a star is concerned, the possible cause for the phenomenon as the interaction of dark matter with the cosmic strings is discussed.  

It can be proposed that the interaction between dark matter and cosmic strings could lead to a dimensional reduction and cut-off of the dimensions around a stellar object, resulting in the object escaping observation.

Dark matter gravitationally interacts with the stellar object. This prediction suggests that the dark matter interaction with the cosmic string will lead to a dimensional reduction.
They will be the result of any gravitational interaction as dark matter couples gravitationally with spherically symmetric stellar objects \cite{tucker1998dark}.
Such gravitational interactions of dark matter and cosmic strings will lead to the cut-off of the dimensions around the stellar object. Hence, the object will escape from the observations. The dark matter clouds too may barrier the observation of a stellar object.
A rare phenomenon happened to Betelgeuse as it started the dimming. The Betelgeuse starts to disappear from October 2019 through February 2020. But currently, it has regained its original brightness. Some speculation on this dimming event is that the Betelgeuse might have been surrounded by a dark matter cloud. The theory suggests that the dark matter candidate axions might have flooded out of Betelgeuse. 
\subsection{Mass to radius }
The mass-to-radius relationship is also an important factor for the evolution and dynamics of the stellar object \cite{demircan1991stellar}. 
\begin{equation}\label{mr1}
\log R=
\begin{cases}
-0.0531+0.8824 \log M, & \text{if}\ M < 1.06 \\
0.0088 + 0.5615 \log M, & \text{if}\ M > 1.47
\end{cases}
\end{equation}
As the mass of the dark matter is added to the stellar object, the total mass and radius parameters will eventually increase. Hence, the stellar object will suddenly disappear due to the interaction between the dark matter and cosmic strings in its trajectory. The sudden increment of curvature and reduction of dimension will cause the star to disappear immediately without any trace. 

 \section{Dark matter and dwarf galaxies}
 Dwarf galaxies are small, faint galaxies that are typically found in the vicinity of larger galaxies like the Milky Way. They are believed to be rich in dark matter, with dark matter making up a large portion of the total mass of the galaxy.
 
 The presence of dark matter around dwarf galaxies can be inferred through a number of observational techniques, such as gravitational lensing and the observation of the motion of stars and gas within the galaxy.
 
 One of the most famous examples of a dwarf galaxy rich in dark matter is the Sagittarius Dwarf Galaxy \cite{jiang2000orbit}. This galaxy is located close to the Milky Way and is thought to contain a large amount of dark matter. Studies have found that the galaxy's stars are moving too fast to be held together by the galaxy's visible mass alone, and that dark matter must be present to explain the observed motion. The Fornax Dwarf Galaxy is another example, for the dark matter present in the dwarf galaxy \cite{strigari2006large}.
 
 It's worth noting that Dwarf Galaxies are known to be rich in the dark matter but the exact distribution and properties of the dark matter within them are not well understood. They are very challenging to study due to their faintness and distance, but they are important objects to study as they can provide clues to the nature and behavior of dark matter.

 One way to study these interactions is by looking at the collisions between dark matter particles and stars. The equations that describe these interactions are based on the principles of quantum mechanics and the weak nuclear force.
 
 The rate of dark matter-star collisions can be calculated using the following equation:
\begin{equation}
 R = n  \sigma  v  A
\end{equation}

 Where $R$ is the collision rate, $n$ is the number density of dark matter particles in the vicinity of the star, $\sigma$ is the cross-section for the collision (i.e. the probability of the collision occurring), $v$ is the relative velocity of the dark matter particle and the star, and $A$ is the area of the star that can be hit by the dark matter particle.
 
 The cross-section for the collision can be calculated using the following equation:
\begin{equation}
 \sigma = \frac{G^2  M^4}  {4 \pi  (E + m)^2}
\end{equation}

 Where $G$ is the weak nuclear force coupling constant, $M$ is the mass of the dark matter particle, $E$ is the energy of the dark matter particle, and $m$ is the mass of the star.

 \section{Stellar evolution}\label{stev}

 The pressure of a homogeneous star is described in terms of the gravitational constant $G$, the mass of the star $(m)$, and the radius of the star $(R)$. 
 
 For the homogeneous star, the pressure is obtained as 
 \begin{equation}\label{st1}
 P \sim \frac{G m^2}{R^4}
 \end{equation}
 For polytrope stars
 \begin{equation}\label{st2}
 \frac{d P}{d r} = - \frac{G m_r}{r^2} \rho
 \end{equation}
 The equation \ref{st2} is describing the relationship between pressure and density for a class of stars known as polytrope stars. These are stars that follow a specific equation of state, known as the polytropic equation of state, which relates the pressure and density of the star in a particular way. In this equation, $P$ is the pressure, $r$ is the radial coordinate, $m_r$ is the mass enclosed within radius $r$, and $\rho$ is the density of the star at that radius. The negative sign in front of the term $Gm_r/r^2$ indicates that the pressure decreases as the radius increases.
 The Poisson equation becomes
 \begin{equation}\label{st3}
 \frac{1}{r^2} \frac{d}{d r} \left(\frac{r^2}{\rho} \frac{d P}{d r}\right) = - 4 \pi G \rho
 \end{equation}
 The equation \ref{st3} shows that the distribution of matter in a system determines the distribution of the gravitational potential energy, and vice versa.

 The polytropic relation between pressure and density is represented as
 \begin{equation}\label{st4}
 P = k \rho^{1+\frac{1}{n}}
 \end{equation}
 in equation \ref{st4} $k,n$ are real and positive constants, and $n$ is the polytropic index.
 
 The equation \ref{st4} the variable $n$ is known as the polytropic index, it is a measure of the degree of compressibility of the gas inside a star. The variable $k$ is a constant of proportionality, it depends on the specific properties of the star.
 
 The polytropic relation is used to model the internal structure of stars, it simplifies the mathematical description of the complex physical processes that occur inside a star. It is a widely used model in the field of stellar astrophysics, and it can be used to estimate the structure and evolution of stars of different sizes and temperatures.
 
 It is important to note that the polytropic relation does not take into account the detailed microphysics of the star, such as the energy transport mechanisms, radiation, and chemical reactions. However, it is a useful tool to understand the overall properties of a star, such as its radius, temperature, and luminosity.
 
 In a large-scale scenario, the effect of dark energy may be dominance. Hence the dynamics are introduced with dark energy. 

 The dominance of dark energy in equation \ref{st4} is implemented as,
 \begin{equation}\label{st5}
 P = e^\lambda k \rho^{1+\frac{1}{n}}
 \end{equation}
 
 In equation \ref{st5}  $e^\lambda$ is the cosmological constant, $k$ and $n$ are constants and $n$ is the polytropic index.
 
 One can define the radial component as 
 \begin{equation}\label{st6}
 r = \alpha \xi 
 \end{equation} 
 with $\xi$ the radius like a variable, and
 \begin{equation}\label{st7}
 \alpha^2 = \frac{k (n+1) \rho_c^{\frac{1-n}{n}}}{4 \pi G}
 \end{equation}
 $\rho_c$ is the central density. 
 
 The equation \ref{st6} is defining a radial variable, denoted by $\xi$, in terms of a dimensionless variable $\alpha$ and the variable $r$.
 
 The variable $\alpha$ is defined in equation \ref{st7} in terms of constants $k$, $n$, and $\rho_c$, as well as the gravitational constant $G$. It can be seen that $\alpha$ has units of length, and so the variable $\xi$ has the same units as $r$.
 
 The central density, $\rho_c$, represents the density of the material at the center of the star, which is an important parameter in determining the properties of the star. The parameter $k$ and n are related to the polytropic index, and they are related to the equation of state of the star, which defines how the pressure and density of the star vary with radius.
 The Poisson equation becomes
 \begin{equation}\label{st8}
 \frac{1}{\xi^2} \frac{d}{d \xi} \left(\xi^2 \dfrac{d \theta}{d \xi}\right) = - \theta^n
 \end{equation}
 
 Equation \ref{st8} is a modified form of the Poisson equation, which is used to describe the gravitational potential in a system.  The equation \ref{st8} is in the form of Lane-Emden equation \cite{adomian1995analytic,van2008analytic}.
 
 Equation \ref{st8} is a dimensionless equation that describes the structure of a self-gravitating, spherically symmetric, non-rotating fluid. The variable $\theta$ is related to the density of the fluid, and the variable $\xi$ is a dimensionless radius.  On the left side of the equation, the first term represents the change in the gravitational potential with respect to the radius and the second term is the change in the density with respect to the radius. On the right side of the equation, the term $-\theta^n$ represents the effect of the fluid's self-gravity on the potential. The value of the parameter $n$ is referred to as the polytropic index and it determines the equation of the state of the fluid. The solution to the Lane-Emden equation gives the distribution of density and pressure throughout the fluid, and it is used to study the structure and evolution of stars.
 
 At polytropic temperatures $\theta = 1$, then the solution is
 \begin{equation}\label{st9}
 \frac{d \theta}{d \xi} = 0
 \end{equation}
 The equation retains at $\xi=0$. The equation \ref{st9} represents the solution for the temperature of a polytrope star when the temperature is at a constant value of $1$.  The solution in equation \ref{st9} implements that at the radial component approaches to minimal values as, like equation \ref{c2.1}, the temperature remains conserved. But the stellar object will hide away from the observations. This analysis in equation \ref{st9} implements the sudden disappearance of the stellar object irrespective of the instant collapse. Regarding these solutions, for the sudden reduction of radial components, the cosmic string solutions are inquired. The nature of the cosmic string while it interacts with the stellar object, let the sudden disappearance of a degree of freedom.

\section{Modified stellar evolution}

Using the solutions and calculations from the above sections, the problem of the mysterious disappearance of the star PHL293B-LBV can be discussed.

The PHL 293B is a metal-poor blue compact dwarf galaxy about 22.6 Mpc from the Earth in the constellation Aquarius \cite{burke2020curious,tenorio2015origin}. It was first listed as entry 293 in a catalog of faint blue stars published by Guillermo Haro and Willem Jacob Luyten in 1962. 
In 1965, Thomas Kinman observed two faint possible companions to it, about $1'$ away, which he dubbed A and B. HL 293B, sometimes called Kinman's Dwarf, was noted to be an extragalactic, nonstellar object, with a jet, approximately 22.6 Mpc away from Earth. The acronym PHL has since been applied to distinguish it from other HL catalogs; it is most commonly referred to by astronomers as PHL 293B. The galaxy was identified as a blue compact dwarf, a type of small irregular galaxy undergoing a strong burst of star formation. Due to its large distance, astronomers could not distinguish the individual stars. In general, such classified stars exhibit shifts in their spectra as well as behave unstably. 

The PHL293B-LBV had  a luminosity as  $ L \ast = 2.5 - 3.5 \times 10^6 L_\odot$, wind velocity as $ 1000 km s^{-1}$, the mass-loss rate as $M = 0.005-0.0200 M_\odot yr^{-1}$  and effective and stellar temperatures of $T_{eff} = 6000 - 6800$  and $T \ast = 9500 - 15000 K$.

The locally coupled dark energy may play a role in the mysterious disappearance of the star PHL293B-LBV. The coupled dark energy will have the strength of nuclear potential. As the term  $X$ gets to infinity, the radial component will also get zero values instantaneously. Hence, the instantaneous disappearance of the stellar object will happen. Dark energy will behave violently on larger cosmic scales. The interaction between dark energy and dark matter will decide the fate of the universe. The dark energy will help the universe evolve without facing future singularities such as big rip \cite{natarajan2020conformal}.

In addition to the dark matter and cosmic string solutions, there exists another possible solution to the local cosmological constant solutions. As the expansion rate is increased, the cosmological constant may have consequences for the dynamics of the stellar object \cite{zubairi2015static}.

The modified TOV equation can be expressed as

\begin{equation}\label{ef1}
\frac{d P}{d r} = - \frac{\epsilon \left(1+\frac{P}{\epsilon}\right) m \left(1 + \frac{4 \pi P r^3}{m} - \frac{\Lambda r^3}{3 m}\right)}{r^2 \left(1-\frac{2 m}{r}-\frac{\Lambda r^3}{3 m}\right)}
\end{equation}

Here $G=c=1$ and $P$ is the pressure, $\epsilon$ is the energy density and $m$ is the mass inside a spherical shell of radius $r$.

This is an equation that describes the behavior of pressure inside a spherical shell of radius $r$, in terms of the energy density, mass and cosmological constant. The modified TOV equation, which stands for Tolman-Oppenheimer-Volkoff, is a modified version of the original equation that describes the behavior of pressure and density inside a spherical shell of radius $r$, in a star. The equation includes the cosmological constant, lambda $(\lambda)$, which is a measure of the density of dark energy in the universe. The equation is derived in the context of the Einstein field equations. The equation is a key tool in understanding the behavior of pressure and density inside stars and is used in the study of stellar structure and evolution.

 From the above results, Equation \ref{ef1} can be related to equation \ref{st3} and satisfied the Poisson equation. Equation \ref{ef1} and equation \ref{st3} are related in the sense that they both describe the equilibrium of a star, but they do so in different ways. Equation \ref{ef1} uses the TOV equation to describe the hydrostatic equilibrium of a star in the presence of a cosmological constant, while equation \ref{st3} uses the Poisson equation to describe the distribution of matter in the star.

The metric is written as 

\begin{equation}\label{ef2}
ds^2 = A  dt^2 - B dr^2 - r^2 d \theta^2 - r^2 sin^ \theta d \phi^2
\end{equation}
The metric in equation \ref{ef2} is the Schwarzschild-de Sitter metric which describes the spacetime around a spherically symmetric object in the presence of a cosmological constant. 
The terms $A$ and $B$ are known as the metric potentials and they determine the curvature of spacetime in the vicinity of the star.
The components of the metric are derived the $A$ and $B= 1/A$ as

\begin{equation}\label{ef3}
A= 1 - \frac{2 m}{r} -\frac{\Lambda r^2}{3}
\end{equation}

The metric in equation \ref{ef2} is the Schwarzschild-de Sitter metric which describes the spacetime around a spherically symmetric object in the presence of a cosmological constant. The term $ds^2$ is the interval of spacetime, $dt$ is the time coordinate, $dr$, $d\theta$, and $d\phi$ are the coordinates for the radial distance, polar angle, and azimuthal angle respectively. The terms $A$ and $B$ are known as the metric potentials and they determine the curvature of spacetime in the vicinity of the star.

In equation \ref{ef3}, the term $m$ is the mass of the star and $\Lambda$ is the cosmological constant. The term $\Lambda r^2/3$ corresponds to the dark energy density, which is assumed to be in the order of nuclear density scales $(140 MeV/fm^3)$  \cite{carroll2001cosmological}. This term is responsible for the acceleration of the expansion of the universe. The term $A$ can be related to the spacetime curvature, and it describes the gravitational pull of the star on the surrounding spacetime.

The Kertsmann scalar corresponds to the dark energy-dominated stellar structure and is represented as

\begin{equation}\label{ef4}
K = R_{abcd} R^{abcd}=\frac{8 \, {\left(\Lambda^{2} r^{6} + 18 \, m^{2}\right)}}{3 \, r^{6}}
\end{equation}

The term $\Lambda^2 r^6$ in the numerator of equation \ref{ef4} corresponds to the dark energy density. The term $18 m^2$ in the numerator corresponds to the mass of the star. The denominator of the equation, $3r^6$, serves as a normalization factor. The ratio of the numerator and denominator of the equation describes the deviation of the spacetime from flatness and gives an idea of how much the spacetime is curved by the dark energy-dominated stellar structure.

The metric satisfies the Einstein field equations as The stress-energy tensor is calculated for the above stellar object with dark energy domination.

\begin{equation}\label{ef5}
\begin{split}
T = \left( \frac{{\left(\Lambda^{2} r^{6} + 12 \, m \Lambda r^{3} - 3 \, \Lambda r^{4} + 36 \, m^{2} - 18 \, m r\right)} p\left(t\right)}{9 \, r^{2}} \right) \mathrm{d} t\otimes \mathrm{d} t \\+ \left( \frac{{\left(\Lambda^{2} r^{6} + 12 \, m \Lambda r^{3} - 6 \, \Lambda r^{4} + 36 \, m^{2} - 36 \, m r + 9 \, r^{2}\right) \rho(t)}} {9 \, r^{2}} \right) \mathrm{d} t\otimes \mathrm{d} t  \\+ \left( -\frac{3 \, r p\left(t\right)}{\Lambda r^{3} + 6 \, m - 3 \, r} \right) \mathrm{d} r\otimes \mathrm{d} r + r^{2} p\left(t\right) \mathrm{d} {\theta}\otimes \mathrm{d} {\theta} \\+ r^{2} p\left(t\right) \sin\left({\theta}\right)^{2} \mathrm{d} {\phi}\otimes \mathrm{d} {\phi}
\end{split}
\end{equation}

Equation \ref{ef5} expresses the stress-energy tensor $T$ of the dark energy-dominated stellar object, where $p(t)$ and $\rho(t)$ are the pressure and density of the star as a function of time, respectively. Here, the stress-energy tensor includes contributions from the dark energy density, which is determined by the cosmological constant $\Lambda$. The equation also relates the pressure, density, and curvature of the spacetime.

\subsection{Case1}
The cases for the $A$ in the metric (\ref{ef2}, \ref{ef3})
are written as

\begin{equation}\label{c1.1}
r = - \frac{2m G}{\left(\frac{\Lambda r^2}{3}-1\right)}
\end{equation}

As $\Lambda \to \infty$ and $r \to 0$, the radial components vanish. This results in the sudden disappearance of the stellar object. The locally incremented cosmological constant is the possible consequence of the sudden incident.

The equation \ref{c1.1} describes the relationship between the radial component of the star, represented by the variable $"r"$, and the cosmological constant, represented by the variable $\Lambda$. The equation shows that as the cosmological constant becomes larger, the radial component of the star becomes smaller. The solution of the equation, as represented by the statement "as $\Lambda \to \infty$ and $r \to 0$", means that as the cosmological constant becomes infinitely large, the radial component of the star becomes infinitely small, resulting in the sudden disappearance of the stellar object. This scenario is caused by the locally incremented cosmological constant, which can be thought of as a possible explanation for the mysterious disappearance of the star PHL293B-LBV.

\subsection{Case2}

The radial component is written as

\begin{equation}\label{c2.1}
r = \frac{2 m G}{1 -X}
\end{equation}

As $X \to \infty$ and $r \to \infty$. That leads to the sudden disappearance of the radial component. But if the same situation is perturbed by cosmic strings, that will cut off the remaining degrees of freedom. The $X$ may be the local interaction of the cosmic string. The $X$ component may be the locally increased curvature due to the cosmic strings. Hence, the sudden increment of curvature will lead to the sudden disappearance of the star. 

In the case of equation \ref{c2.1}, the radial component of the star is inversely proportional to a term denoted as $X$. As $X$ approaches infinity, the radial component of the star approaches zero, which results in the sudden disappearance of the star.  In this case, it is suggested that the presence of a cosmic string locally increases the curvature of spacetime, causing the radial component of the star to decrease rapidly and leading to its sudden disappearance.

Hence, we can conclude 
\begin{equation}\label{c2.3}
ds^2=r=\phi =\theta=0
\end{equation}

Equation \ref{c2.3} is a mathematical representation of the sudden disappearance of the radial component, the angle $\phi$, and the angle $\theta$ of the stellar object.

By combining the discussed possible solutions, one can predict that the local domination of the cosmological constant, or the interaction of cosmic strings may result in its sudden reduction of the radial component. As the consequence, the stellar object vanishes from the observations, which can be a possible reason for  the sudden disappearance of the star PHL293B-LBV.

 \section{Simulation Results}

 To predict the correctness of the results, the IllustrisTNG simulation is considered. \cite{illu01}. 
 
 The Illustris simulations are based on the latest understanding of cosmology and the physical processes that govern the formation and evolution of structure in the universe. The IllustrisTNG simulations also provide valuable insights into the large-scale structure of the universe, such as the distribution of galaxy clusters and distribution of dark matter. They are also used to make predictions about the properties of the universe that can be tested against observational data, such as the cosmic microwave background and large-scale structure.

 The column density of dark matter is a measure of the amount of dark matter present along a line of sight through a given region of space. It is typically measured in units of mass per unit area, such as grams per square centimeter. The column density is calculated by integrating the local density of dark matter along the line of sight. Observations of the gravitational lensing effect can also be used to infer the column density of dark matter.

\begin{figure}
	\includegraphics{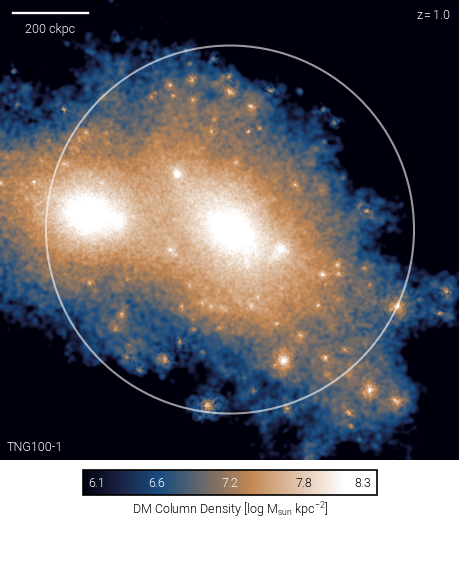}
	\caption{Simulation of dark matter column density}
	\label{fig:tng}
\end{figure}

 In image \ref{fig:tng} the simulation of dark matter column density is reported. With the configurations of $200 ckpc$ and $z=1.0$, the dark matter column density has a greater magnitude in the core of this region. From the edges, the density parameter increases towards the core. Also, There is another region where a similar kind of increased dark matter column density.  These simulations show that the presence of dark matter density in the local galaxies. Hence this result can support the role of dark energy in the mysterious disappearance of a local star which is a consequence of the direct or indirect effect of dark matter. 
 
 \begin{figure}
 	\includegraphics{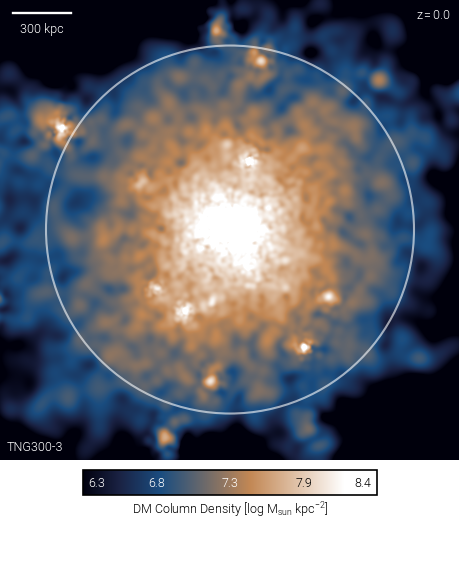}
 	\caption{Dark matter column density at the center of the dwarf galaxy}
 	\label{tng2}
 \end{figure}

Similarly, for the case of large-scale universe structures, simulations from TNG300 were reported for the dark matter column density. The simulation provides detailed information about the dark matter column density, which is a measure of the amount of dark matter present in a given area. The image referred to as "image \ref{tng2}" is a visual representation of the simulation's results. The image shows the distribution of dark matter in a specific region of the universe, and it is clear that the density of dark matter increases towards the core of the selected region.

These simulations are consistent with what is expected from the observed distribution of dark matter in the universe. Dark matter is thought to be more concentrated in the cores of galaxy clusters and superclusters, where the gravitational pull is stronger. Therefore, the simulation's results support the proposed presence of dark matter in local clusters and suggest that the TNG300 project is capable of providing accurate and detailed information about the distribution of dark matter in the universe.

\subsection{Alternative possiblities}

 There are a few other possible explanations for why a star might suddenly disappear from observations. One possibility is that the star has gone supernova, which is a catastrophic explosion that occurs when a massive star runs out of fuel and collapses. When a star goes supernova, it can release a tremendous amount of energy, and the explosion can be so bright that it is visible even during the day. However, after the explosion, the star itself is usually no longer visible because it has been destroyed or has collapsed into a compact object like a neutron star or black hole.
Another possibility is that the star has simply moved out of view. Stars are constantly in motion, and it is possible that a star that was previously visible from Earth has moved behind another object in the sky, such as a planet or another star. In this case, the star might still be there, but it is simply not visible from our perspective.
Finally, it is also possible that the star has simply dimmed or faded. Some stars are prone to fluctuating in brightness, and it is possible that a star that was previously bright has suddenly become dimmer or has gone into a period of inactivity. This could be due to various factors, such as the star running out of fuel or experiencing changes in its internal structure.

Additionally, the idea of stars disappearing into extra dimensions is a theoretical concept that has been proposed as a possible explanation for certain observed phenomena in the universe, such as the sudden disappearance of stars or the lack of certain types of objects in certain parts of the universe. According to some theories, our universe may have more than three spatial dimensions, and the extra dimensions may be "compactified" or "curled up" on a very small scale, such that they are not directly observable. It is possible that some objects, such as stars, may have the ability to "fall into" these extra dimensions, effectively disappearing from our observable universe. This idea is based on the concept of brane cosmology, which suggests that our universe may be a three-dimensional "brane" embedded in a higher-dimensional "bulk" space. In this scenario, it is possible that some objects may be able to move between the brane and the bulk, effectively disappearing from the brane and reappearing in the bulk.
 
 \section{Conclusion}

For the case of the mysterious disappearance of the star,  traditional explanations such as conversion to a stellar black hole, dust gases, and Dyson spheres have been ruled out due to a lack of observational evidence.
 
 Our proposed model suggests that the disappearance of the massive star PHL293B-LBV may be the result of the interaction of dark matter with a cosmic string or the dark energy field from higher dimensions. The recent release of Gaia datasets \cite{gaia2022d}  has provided information about the positions, distances, and dynamics of various stellar objects.  The Gaia dataset  contains information about 2 billion stars, solar system objects, and extragalactic sources has revealed that stars undergo massive tidal forces, known as "star tsunamis," which can cause changes in the shape and size of stars.  We propose that such tidal forces, as a consequence of external interactions, may lead to the sudden disappearance of stars without any observable impacts. In the future, such events may be captured via gravitational wave observations by installing observatories in outer space.
 
 However, it is important to note that the proposed model is based on theoretical assumptions and requires further investigation through observational data and numerical simulations.  The sudden disappearance of PHL293B-LBV presents a unique opportunity for further research and understanding of the complex interplay between dark matter, cosmic strings, and dark energy in the universe.

\bibliography{ref}
\bibliographystyle{plain}

\end{document}